\documentclass[aps,pra,reprint,showpacs,floatfix]{revtex4-1}

\usepackage{amsmath}
\usepackage{slashed}
\usepackage{txfonts}
\usepackage{microtype}
\usepackage{graphicx}
\usepackage[
breaklinks=true
]{hyperref}
\usepackage{color}
\usepackage{isomath}
\def\sout{\bgroup\markoverwith
{\textcolor{red}{\rule[0.5ex]{2pt}{0.5pt}}}\ULon}
\def\be{\begin{equation}}
\def\ee{\end{equation}}
\def\bes{\begin{equation*}}
\def\ees{\end{equation*}}
\def\bea{\begin{eqnarray}}
\def\eea{\end{eqnarray}}
\def\beas{\begin{eqnarray*}}
\def\eeas{\end{eqnarray*}}
\def\bal#1\eal{\begin{align}#1\end{align}}
\def\bals#1\eals{\begin{align*}#1\end{align*}}
\newcommand{\bra}[1]{\langle #1|}
\newcommand{\ket}[1]{|#1\rangle}
\newcommand{\braket}[2]{\langle #1|#2\rangle}

\renewcommand{\vec}[1]{\mathbf{#1}} 
\newcommand{\del}{\partial}

\graphicspath{{figures/}}

\renewcommand*{\vec}[1]{\boldsymbol{#1}}

\begin{document}
\title{The propagator of a relativistic particle via the path-dependent vector 
potential}

\author{Enderalp \surname{ Yakaboylu}} 
\affiliation{Max-Planck-Institut f\"ur Kernphysik, Saupfercheckweg 1, D-69117 
Heidelberg, Germany}
\author{Karen Z. \surname{Hatsagortsyan}} 
\affiliation{Max-Planck-Institut f\"ur Kernphysik, Saupfercheckweg 1, D-69117 
Heidelberg, Germany}
\author{Christoph H. \surname{Keitel}}
\affiliation{Max-Planck-Institut f\"ur Kernphysik, Saupfercheckweg 1, D-69117 
Heidelberg, Germany}
\date{\today}

\begin{abstract}

The proper time formalism for a particle propagator in an external 
electromagnetic field is combined with the path-dependent formulation of the 
gauge theory to simplify the quasiclassical propagator. The latter is achieved due to a specific
choice of the gauge corresponding to the use of the classical path in the path-dependent
formulation of the gauge theory, which leads to the cancellation of the 
interaction part of the action in the Feynman path integral. A simple expression 
for the quasiclassical propagator is obtained in all cases of the external field 
when the classical equation of motion in this field is integrable. As an 
example, new simple expressions for the propagators are derived for a spinless 
charged particle interacting with the following fields: an arbitrary constant 
and uniform electromagnetic field, an arbitrary plane wave and, finally, an 
arbitrary plane wave combined with an arbitrary constant and uniform 
electromagnetic field. In all these cases the quasiclassical propagator 
coincides with the exact result.

\end{abstract}

\pacs{03.65.Pm, 03.65.Db, 03.65.Sq}

\maketitle

\section{Introduction}

For those potentials which vary much more slowly than the corresponding wave function in
position, the wave function given within the Wentzel-Kramers-Brilluin (WKB) approximation becomes
legitimate \cite{Weinberg_qm}. In a sense, we are in the quasiclassical regime when the de Broglie
wavelength of the particle varies slightly over the distance that characterizes the problem
\cite{Landau_3}. In fact, if the Lagrangian of a point particle is a quadratic function of the
coordinate and the velocity, then the corresponding exact propagator
coincides with the quasiclassical propagator, i.e., the classical path dominates the 
Feynman path integral \cite{Feynman_1948,Feynman_Hibbs,Schulman,Kleinert}. It 
is interesting that the exact Volkov propagator, the propagator of a charged 
particle interacting with a plane electromagnetic  wave, is also given
by its quasiclassical limit \cite{Volkov_1935, 
ReissEberly,EberlyReiss,Boudjedaa_1992, Barducci_2003}, though the Lagrangian
is not quadratic. Furthermore, there are many cases where the quasiclassical 
propagator is a good approximation. In this paper, we show that the expression 
of the quasiclassical propagator can be significantly simplified when employing 
the path dependent formulation of the vector potential within the proper time 
formalism for calculation of propagators. 

The path-dependent formulation of the gauge theory is developed by DeWitt and 
Mandelstam in order to discuss quantum theory without electromagnetic potentials
\cite{DeWitt_1962,Mandelstam_1962}. 
This formalism emerges from the nonintegrable phase factor
which describes the complete electromagnetism as shown by Wu and Yang 
\cite{Wu_Yang_1975,Yakaboylu_2013}. In this equivalent formulation of the gauge 
theory the vector potentials are defined via a path-integral over 
electromagnetic fields such that each path corresponds to a 
certain gauge function 
\cite{Belinfante_1962,Rohrlich_Strocchi_1965,Yakaboylu_2013}, 
i.e., the gauge is determined by the choice of the path in the corresponding 
integral, rather than by the gauge function as in the common gauge theory. The 
formalism further provides a geometric picture for the gauge transformation, 
expressing the gauge function via the electromagnetic flux. 

The propagator of a particle in an electromagnetic field can be calculated via 
the Feynman path integral \cite{Feynman_1948,Feynman_Hibbs}, 
where the propagator is expressed in terms of the action. When additionally the 
path-dependent formulation of the gauge representation is used, 
the interaction part of the action of the Feynman path integral
can be expressed in terms of the electromagnetic flux through the area between 
the arbitrary Feynman path and the gauge path which generates the associated 
path-dependent vector potential \cite{Yakaboylu_2013}. 
The significant simplification of the quasiclassical propagator expression 
comes when one specifies the gauge path to coincide with the classical 
trajectory. In fact, as we will show here, the above mentioned flux vanishes in 
this case and, consequently, the interaction part of the action vanishes in the 
Feynman path integral.

The straightforward calculation of the relativistic propagator via the 
Feynman path integral is cumbersome due to the presence of the particle's 
infinitesimal proper time \cite{Feynman_1948,Feynman_Hibbs,Schulman,Kleinert}. 
However, one can overcome this difficulty via introducing a fifth parameter to 
the theory which is known as the proper time formalism 
\cite{Fock_1937, 
Feynman_1950,Schwinger_1951,Dirac_qm,Fradkin_1991,Brink_1977,Polyakov_st}. In 
this method one defines an effective Lagrangian associated to the 
super-Hamiltonian $\mathcal{H}\left( P, X \right)$.
The latter is defined by the quantum mechanical equation: any 
quantum mechanical equation can be written in the form of 
\be
\label{qm_eqn}
\mathcal{H} \left( P, X \right) \,  \ket{\psi} = 0,
\ee
with the four-momentum operator $P$ and four-position operator $X$. Then, the 
relativistic propagator is followed by the standard rules of the nonrelativistic 
Feynman path integral for an effective system governed by the effective 
Lagrangian.

In this paper, we unite two powerful methods for the calculation of the propagator 
of a relativistic charged particle interacting with an external electromagnetic 
field. In particular, the path-dependent formulation of the gauge theory is 
incorporated in the proper time formalism of propagators.  
This allows us to obtain simple expressions of the quasiclassical propagators
for a constant and uniform electromagnetic field, for an arbitrary plane wave and
for an arbitrary plane wave combined with an arbitrary constant and uniform electromagnetic
field. In all these cases, the quasiclassical propagator
coincides with the exact result.

The structure of the paper is the following. In Sec.~\ref{PDF},
the path-dependent formulation of gauge theory is presented and
its convenience for the Feynman path integral method for calculation of 
propagators is shown. In the next section, Sec.~\ref{PT}, the proper time 
formalism for calculation of propagators is introduced  
which incorporates the path-dependent gauge formalism. The explicit expressions 
of the propagators for a spinless charged particle interacting with a constant 
and uniform electromagnetic field, with a plane wave and with a plane wave 
combined with a constant and uniform electromagnetic field, respectively, are 
derived in Sec.~\ref{app}. The conclusion is given in Sec.~\ref{conc}. 

The metric convention is $g = (+,-,-,-)$ throughout the paper.

\section{The path-dependent formalism of the gauge theory} \label{PDF}

The  Maxwell equations allow to express the electromagnetic field strength tensor $F^{\mu \nu}$ 
 in terms of a four-vector potential $A^\mu = (\phi, \vec{A})$ as
$F_{\mu \nu}(x) =  \del_\mu A_\nu(x) - \del_\nu A_\mu(x)\, $. The gauge 
transformation
\be
\label{gauge_trans}
A^\mu (x) \rightarrow A^\mu (x) + \del^\mu \chi (x)
\ee
leaves the electromagnetic field strength tensor invariant, which is a 
consequence of the conservation of the electric charge under the local symmetry 
transformation via Noether's theorem. The conservation of the electric charge 
follows from the local phase invariance of the wave function
such that the Schr\"{o}dinger equation is
invariant under the transformations Eq. (\ref{gauge_trans}) as long as the
wave function transforms as
\be
\psi(x) \rightarrow \exp\left(\dfrac{i e \chi(x)}{\hbar c}\right)
\psi(x)\, .
\ee
In fact, the gauge function $\chi$ can be identified via the path integral 
\be 
\chi(x) = - \int_{\mathcal{P}}^x A_\nu(y) d y^\nu .
\ee
This path-dependent phase (nonintegrable phase \cite{Wu_Yang_1975}),
then, yields a gauge invariant but path-dependent vector potential 
$\mathcal{A}_\mu(x)$ as
\be
\label{path_dependent_potential}
\mathcal{A}_\mu(x) = \int_{0}^{1}  F_{\nu \lambda}(y) \dfrac{\del
y^\nu}{\del s}  \dfrac{\del y^\lambda}{\del x^\mu}  d  s
\ee
with the boundary conditions $y(1, x) = x\, , \quad
y(0, x) = x' \, $, where the vector potential vanishes at $x'$. In this 
equivalent formulation of the gauge theory the Schr\"{o}dinger equation is 
invariant under path transformations
\bal
\label{path_transformation}
\mathcal{A}_\mu (\mathcal{P'},x) &= \mathcal{A}_\mu (\mathcal{P},x) + \del_\mu
\Phi_{EM}(\Sigma, x) \, , \\ 
\Psi[\mathcal{P'},x] & = \exp\left(\dfrac{i e}{\hbar c} \Phi_{EM}(\Sigma, x) 
\right)
\Psi[\mathcal{P},x]\, ,
\eal
with the electromagnetic flux
\be
\label{wilson_loop}
\Phi_{EM}(\Sigma, x) = \oint_{\del \Sigma}^x A^\mu d y_\mu = \dfrac{1}{2} 
\int_\Sigma^x F^{\mu \nu}
d \sigma_{\mu \nu}
\ee 
on the closed loop ${\del \Sigma} = \mathcal{P}-\mathcal{P'}$ \cite{Yakaboylu_2013}.
The electromagnetic flux can further be identified as
\be
\label{flux_via_open_int}
\Phi_{EM}(\Sigma,x) = \int_\mathcal{P}^x \mathcal{A}_\mu (\mathcal{P'},y) d 
y^\mu \, ,
\ee
which implies that the line integral of a path dependent vector potential along 
a path $\mathcal{P}$ vanishes as long as the vector potential is evaluated via 
the same path $\mathcal{P}$, i.e., 
\be
\label{vanishing_path_integral}
\int_\mathcal{P}^x \mathcal{A}_\mu (\mathcal{P},y) d y^\mu = 0 \, .
\ee

The full machinery of the path-dependent formulation of the gauge theory provides 
some fundamental simplifications for the path integral formulation of quantum 
mechanics. Namely, let us consider the propagator 
in terms of the Feynman path integral which is defined by 
\be
K_F (\vec{x},\vec{x}';t) = \int D(\mathcal{P}_F) \, \exp\left( \frac{i}{\hbar} 
S(\mathcal{P}_F) \right)
\ee
where $D(\mathcal{P}_F)$ and $S(\mathcal{P}_F)$ represent the sum over all paths 
and the action evaluated along the path  $\mathcal{P}_F$, respectively. The 
action of interest of the present manuscript is the action of a spinless 
charged particle interacting with an electromagnetic field, which can be written 
as
\be
\label{em_action}
S(\mathcal{P}_F) = - mc^2 \int_{\mathcal{P}_F}  d \tau - \dfrac{e}{c} 
\int_{\mathcal{P}_F}
\mathcal{A}_\mu (\mathcal{P}_G,y) dy^\mu \, ,
\ee
with the particle's infinitesimal proper time $d \tau = \sqrt{d y^\mu d y_\mu}/c 
$. Here we should emphasize that paths appeared in the Feynman path integrals 
$\mathcal{P}_F$ are real paths in the sense that the transition amplitude of a 
particle from spacetime point $x'$ to $x$ depends on these paths. Nonetheless, 
paths used for the vector potential $\mathcal{P}_G$ are
just gauge paths. Moreover, the action (\ref{em_action}) can
be defined via the electromagnetic flux~(\ref{flux_via_open_int}) as
\be
\label{action_with_flux}
S(\mathcal{P}_F) = - mc^2 \int_{\mathcal{P}_F}  d \tau - \dfrac{e}{c} 
\Phi_{EM}(\Sigma, x),
\ee
with ${\del \Sigma} = \mathcal{P}_F-\mathcal{P}_G$.

The compact form of the action (\ref{action_with_flux}) provides us a further 
simplification for the quasiclassical propagators.
The quasiclassical propagator can be defined via the classical action $S_c$, which 
is the action evaluated along the classical trajectory (world line) 
$\mathcal{P}_c $. Furthermore, using the VanVleck-Pauli-Morette formula 
\cite{VanVleck_1928,Morette_1951,DeWitt_1957,Pauli_1973}, it reads
\bal
\nonumber & K_F (\vec{x},\vec{x}';t)) = \sqrt{\left(\frac{1}{2 \pi i 
\hbar}\right)^3 
\det\left(\frac{-\del^2
S_c}{\del
\vec{x} \del \vec{x}'} \right)} \\
& \times \exp\left( - \frac{i}{\hbar} mc^2 \int_{\mathcal{P}_c}  d \tau -
\frac{i}{\hbar} \dfrac{e}{c} \Phi_{EM}(\mathcal{P}_c -\mathcal{P}_G, x) \right) 
\, . 
\eal
Now, if one specifies the classical path for the gauge path, the flux term in 
the above expression vanishes and then the quasiclassical propagator reduces to
\bal
\label{WKB}
 & K_F (\vec{x},\vec{x}';t) = \sqrt{\left(\frac{1}{2 \pi i \hbar}\right)^3 
\det\left(\frac{-\del^2
S_c}{\del
\vec{x} \del \vec{x}'} \right)} \exp\left( - \frac{i}{\hbar} mc^2 
\int_{\mathcal{P}_c} \, d
\tau \right) \,,
\eal
where the classical path $\mathcal{P}_c$ satisfies the Lorentz force law
\be
\label{lorentz_force_law}
m \dfrac{\del^2 y_c^\mu}{\del \tau^2} = \dfrac{e}{c} {F^\mu}^\nu (y_c) \dfrac{\del
{y_c}_\nu}{\del \tau} \, .
\ee
The latter defines the path dependent vector potential 
(\ref{path_dependent_potential}) as
\be
\label{vector_potentail_in_pdf}
\mathcal{A}_\mu (\mathcal{P}_c, x) = -\dfrac{m c}{e} \int_{\tau_i}^{\tau_f}  
\dfrac{\del^2
{y_c}_\nu}{\del \tau^2}
\dfrac{\del y_c^\nu}{\del x^\mu} \, d  \tau \, ,
\ee
where the dependence on the electromagnetic fields contains only in the 
definition of the classical path via Eq.~(\ref{lorentz_force_law}). Furthermore, in terms of the
reparametrization invariant form of the equations of motion
\be
\label{lorentz_force_law_par_inv}
\dfrac{\del p^\mu}{\del s} = \dfrac{e}{c} {F^\mu}^\nu (y_c) \dfrac{\del
{y_c}_\nu}{\del s} \,
\ee
with the particle's four momentum $p^\mu$, the vector potential for the classical path reads
\be
\label{vector_potentail_in_ct_par_inv}
\mathcal{A}_\mu (\mathcal{P}_c, x) = -\dfrac{c}{e} \int_{0}^{1}  
\dfrac{\del p_\nu}{\del s} \dfrac{\del y_c^\nu}{\del x^\mu} \, d  s \, .
\ee

For a nonrelativistic particle, on the other hand, the classical trajectory can be parametrized with
the physical time $t'$. Then, the path dependent vector potential for the nonrelativistic classical
path $\vec{y}(t')$ with the boundary conditions $\vec{y}(0) = \vec{x}'$ and 
$\vec{y}(t) = \vec{x}$ can be written as
\be
\label{vector_potentail_for_nr_ct}
\mathcal{A}_\mu (\mathcal{P}_c, x) = \dfrac{m c}{e} \int_{0}^{t} 
\dfrac{\del^2 \vec{y}}{\del t'^2} \cdot \dfrac{\del \vec{y}_c}{\del x^\mu} \, d t'= \dfrac{c}{e}
\int_{0}^{t} 
\dfrac{\del \vec{p}}{\del t'} \cdot \dfrac{\del \vec{y}_c}{\del x^\mu} \, d t' \, ,
\ee
where we have used $\del t' / \del x^\mu = 0 $. This form of the vector
potential~(\ref{vector_potentail_for_nr_ct}) provides great convenience for the quasiclassical
propagator for a nonrelativistic particle.  Since the integral of the corresponding potential terms
in the action vanishes, i.e.,
\be
\int_0^t d t' \, \left( \frac{e}{c} \vec{\mathcal{A}} \cdot \dot{\vec{y}}(t') - e \mathcal{A}^0
\right) = 0 \, ,
\ee
the quasiclassical propagator for a nonrelativistic particle in the classical path gauge yields
\bal
\label{WKB_nonrel}
 & K_F (\vec{x},\vec{x}';t) = \sqrt{\left(\frac{1}{2 \pi i \hbar}\right)^3 
\det\left(\frac{-\del^2
S_c}{\del
\vec{x} \del \vec{x}'} \right)} \exp\left(\frac{i}{\hbar} 
\int_{\mathcal{P}_c}^t 
\frac{ m \, \dot{\vec{y}_c}^2 }{2} d t' \right) \, .
\eal
In fact, this result can also be obtained directly via approximating the particle's infinitesimal
proper time with the usual nonrelativistic kinetic term of the Lagrangian in 
Eq.~(\ref{WKB}). 

In summary, since the equations of motion are gauge invariant, they are 
first found in any convenient gauge, then the propagator for a nonrelativistic 
particle as well as for a relativistic particle can be calculated in 
the corresponding classical path gauge via Eq.~(\ref{WKB_nonrel}) and 
Eq.(\ref{WKB}), respectively. The compact form of the quasiclassical propagator 
can be applied any type of potential, when the classical equations of motion 
are known.

\section{The Proper Time Formalism and The Feynman Kernel via the path dependent 
vector potential} \label{PT}

The calculation of the relativistic propagator is, in general, tedious due to 
the presence of the particle's infinitesimal proper time. Nevertheless, this can 
be overcome via the proper time (eigentime, fifth parameter or einbein) 
formalism  \cite{Fock_1937, 
Feynman_1950,Schwinger_1951,Dirac_qm,Fradkin_1991,Brink_1977,Polyakov_st}.

The proper time formalism is based on the fact that any quantum
mechanical equation can be written in the form of
\be
\label{qm_eqn}
\mathcal{H} \left( P, X \right) \,  \ket{\psi} = 0 \, ,
\ee
with a certain operator $\mathcal{H}(P,X)$, so called super-Hamiltonian. For instance, in the case of Klein-Gordon equation 
the super-Hamiltonian is
\be
\label{KG}
\mathcal{H} = \left(P - \frac{e}{c}\mathcal{A}(\mathcal{P}_G, X) 
\right)^2 -  m^2 c^2 \, .
\ee
The four-momentum operator $P^\mu$ 
and the four-position operator $X^\mu$ in Eq. (\ref{qm_eqn}) satisfy the 
eigenvalue equations
\bal
P^\mu \ket{p} &= p^\mu \ket{p} , \\
X^\mu \ket{x} & =  x^\mu \ket{x},
\eal
respectively, with $p^\mu = (E/c , \vec{p})$ and $x^\mu = (c \, t, \vec{x} )$. 
Further, the assumption of the canonical commutation relation $\left[ X^\mu, 
P^\nu \right]~=~i~\hbar~\,g^{\mu \nu}$ implies
the relations $ \bra{x} P_\mu \ket{\psi} = i \hbar \del_\mu \psi(x)$, $ \bra{x} 
X_\mu \ket{\psi} =
x_\mu \psi(x)$,
\bal
\braket{x}{p} &= \frac{1}{\left( 2 \pi \hbar \right)^2 } \exp\left[ - \frac{i x 
\cdot p }{\hbar}
\right] \, , \\
\braket{x}{x'} &= \delta^4 (x-x') = \int \frac{d^4 p}{(2 \pi \hbar)^4}  
\exp\left[ - \frac{i p \cdot
(x-x') }{\hbar} \right] \, .
\eal
In position space, the fundamental equation~(\ref{qm_eqn}) yields
\be
\bra{x} \mathcal{H} \left(P,X \right) \ket{\psi} = \mathcal{H} \left(i \hbar \, 
\del, x \right) \,
\psi(x) = 0 \, .
\ee
Then, the corresponding Green's function satisfies
\be
\mathcal{H} \left(i \hbar \, \del, x \right) \, G(x,x') = \delta^4 (x-x') \, .
\ee
Hence, the Green's function can be identified as
\be
G(x,x') = \mathcal{H}^{-1} \left(i \hbar \, \del, x \right) \delta^4 (x-x') \,
\ee
which can further read
\be
\label{greens_function_1}
G(x,x') = \bra{x} \mathcal{H}^{-1} \left(P, X \right) \ket{x'} = \bra{x}
\frac{1}{\mathcal{H} \left(P, X \right) + i \epsilon } \ket{x'} \, 
\ee
with the Feynman $ i \epsilon$ prescription. Furthermore, 
Eq.~(\ref{greens_function_1}) may also be defined
as
\be
\label{fundamental_eqn_greens}
G(x,x') = -\frac{i}{\hbar} \alpha \int_0^\infty d \tau   \, \bra{x} \exp\left[ 
\frac{i}{\hbar} \mathcal{H} \left(P, X \right) \alpha \,  \tau 
\right] \ket{x'} e^{-\epsilon \tau} \, 
\ee
with an auxiliary field $\alpha$ (einbein field). 
Here the parameter $\alpha$ is introduced in order to fix the right classical 
equations of motion of the corresponding super-Hamiltonian $\mathcal{H}(P,X)$ 
\cite{Brink_1977,Polyakov_st}.

Let us define the integrand in Eq.~(\ref{fundamental_eqn_greens}) as a Feynman Kernel 
\be
K_F (x,x' ; \tau) = \bra{x}\exp\left[\frac{i}{\hbar} \mathcal{H}
\left(P, X \right) \alpha \, \tau \right] \ket{x'},
\ee
with the effective Hamiltonian $\alpha \mathcal{H}$ in the space of $\tau$, 
which determines the propagator
\be
\label{Green}
G(x,x') = -\frac{i}{\hbar} \alpha \int_0^\infty d \tau \, K_F (x,x' ; \tau) 
e^{-\epsilon \tau} \, .
\ee

The Feynman Kernel in terms of the path integration reads
\be
K_F (x,x' ; \tau) = \int D(\mathcal{P}_F) \, \exp\left( - \frac{i}{\hbar} 
S(\mathcal{P}_F) \right) \, ,
\ee
with the action $S(\mathcal{P}_F)$, which is derived introducing an effective 
Lagrangian via common Legendre transformation
\be
\label{legendre_trans}
\mathcal{L} = p \cdot \dot{x}  - \alpha \mathcal{H} \,
\ee
with $p^\mu = \del \mathcal{L} / \del \dot{x}_\mu$. Then,
\be
K_F (x,x' ; \tau) = \int D[y] \exp\left(- \frac{i}{\hbar} \int_0^\tau d \sigma
\mathcal{L}(y,\dot{y}) \right) \, ,
\ee
where the parametrized path $y(\sigma)$ satisfies the boundary conditions $y^\mu (0) 
= x'^\mu$, $y (\tau) = x^\mu$. 

In the present manuscript, we restrict ourselves to spinless charged particles, 
although the results can be easily generalized to the Fermionic particles. Then 
using Eq.~(\ref{legendre_trans}) for Eq.~(\ref{KG}), the corresponding 
effective Lagrangian is found
\be
\label{lagrangian_of_em}
\mathcal{L} = \frac{1}{4 \alpha} \dot{y}^2 + \frac{e}{c} \dot{y} 
\mathcal{A}(\mathcal{P}_G, y) + \alpha \, m^2 c^2 \, .
\ee
The Euler-Lagrange equation of the effective Lagrangian provides the effective 
equation of motion
\be
\label{effective_lorentz_force}
\frac{1}{2 \alpha}{\ddot{y}(\sigma)_c}^\mu = \frac{e}{c} {F^\mu}_\nu (y_c) 
{\dot{y}(\sigma)_c}^\nu  \, 
\ee
which coincides with the classical equation of motion in the given external 
electromagnetic field for $\alpha = 1 /(2m)$ as long as the 
path is parametrized with the particle's proper time $\tau$, 
see \cite{Zwiebach_st}.  In the 
quasiclassical approximation and using the gauge determined by the classical 
path, see Eq. (\ref{WKB}), the Feynman Kernel yields
\bal
\label{fundamental_feynman_kernel}
\nonumber & K_F (x,x') = \sqrt{\left(\frac{1}{2 \pi i \hbar}\right)^4
\det\left(\frac{\del^2 S_c}{\del
x_\mu \del {x'}_\nu } \right)} \\
& \times \exp\left(- \frac{i}{\hbar} m^2 c^2 \alpha \tau -
\frac{i}{\hbar}
\int_0^\tau d \sigma \, \frac{\dot{y}_c^2}{4 \alpha}  \right) \, .
\eal
The Green function is, then, obtained via Eq.~(\ref{Green}).

The equation (\ref{fundamental_feynman_kernel}) is the main result of the paper, 
which shows how to derive the fundamental Feynman Kernel for the quasiclassical 
Green's function in a simple way when the classical equation of motion 
Eq.~(\ref{effective_lorentz_force}) is integrable in the given field.

\section{Examples} \label{app}

In this section we will apply the developed formalism to some important cases where the
quasiclassical propagator coincides with the exact propagator.

\subsection{Constant and uniform electromagnetic field case}

Let us first consider a spinless charged particle interacting with a 
constant and uniform electromagnetic field $F_{\mu \nu}$. Since the Lagrangian of a constant and
uniform electromagnetic field (\ref{lagrangian_of_em}) is a quadratic function of $y$ and
$\dot{y}$, the quasiclassical formula gives the exact result.

The effective equation of motion Eq. (\ref{effective_lorentz_force}) in the constant and
uniform electromagnetic field is solved providing the classical path 
\be
\label{classical_path_em}
{y(\sigma)_c}^\mu = {\left(\frac{e^{\lambda F \sigma}-1}{e^{\lambda F \tau}-1} 
\right)^\mu}_\nu
(x-x')^\nu + x'^\mu \,
\ee
with $\lambda = 2 \alpha e /c $, and corresponding to the boundary conditions 
${y(0)_c}^\mu= x'^\mu$, ${y(\tau)_c}^\mu=x^\mu$. Here it should be understood 
that
\bal
{\left(\frac{e^{\lambda F \sigma}-1}{e^{\lambda F \tau}-1} 
\right)^\mu}_\nu = 
\frac{\sigma}{\tau}
{g^\mu}_\nu + \frac{\lambda \left(\sigma^2-\sigma \tau \right)}{2 \tau 
}{F^\mu}_\nu + \cdots
\eal

As a result, in the classical path gauge the Feynman Kernel of Eq. (\ref{fundamental_feynman_kernel}) reads
\begin{widetext}
\be
\label{feynman_kernel_classical_path}
 K_F (x,x' ; \tau) = \sqrt{\frac{\det\left[- \frac{\lambda^2 F^2 
\tau} {8 \alpha} \sinh^{-2} (\frac{\lambda F
\tau}{2}) \right]}{(2 \pi i \hbar)^4}} \exp\left(- \frac{i}{4 \alpha \hbar} 
(x-x')\left( \frac{\lambda^2 F^2
\tau}{4} \sinh^{-2} (\frac{\lambda F \tau}{2}) \right) (x-x') - \frac{i m^2 
c^2 \alpha \tau}{\hbar} \right) \, .
\ee
\end{widetext}

The Kernel in an arbitrary gauge, then, is given by the transformation 
\be
\label{gauge_trans_kernel}
K_F (x,x' ; \tau) \rightarrow  \exp\left(-\frac{i e}{\hbar c} \Phi_{EM}(x) 
\right) K_F (x,x'
; \tau) \exp\left(\frac{i e}{\hbar c}\Phi_{EM}(x')\right)
\, 
\ee
where the electromagnetic flux $\Phi_{EM}$ is calculated for the area bounded by 
the loop $\del \Sigma = \mathcal{P}_c - \mathcal{P}_G$ with desired gauge 
$\mathcal{P}_G$. 

Equivalently, the Kernel in an arbitrary gauge can also be 
found using the corresponding gauge function $\chi$, taking into account that
the vector potential in the classical path gauge is
\be
\mathcal{A}_\mu (\mathcal{P}_c, x) = - \frac{1}{\lambda} \int_0^\tau d \sigma 
\, 
\frac{\del^2
{y_c}^\nu}{\del \sigma^2} \frac{\del {y_c}_\nu}{\del x^ \mu}
\ee
where we have used Eq.~(\ref{effective_lorentz_force}) in 
Eq.~(\ref{path_dependent_potential}).

The constant and uniform crossed field with equal amplitude 
\bal
\label{crossed_fields} 
 \vec{E} = E_0 \, \hat{\vec{x}} \, ,\,\,\,\,\,
\vec{B}  = E_0 \, \hat{\vec{y}} \, ,
\eal
is an important particular case of the considered field, corresponding to the 
asymptotic of the laser field at large field parameters $a_0\equiv eA/mc^2\gg 
1$. Since the third power of the field tensor vanishes for the above field, the 
Feynman Kernel in the classical path gauge yields
\bal
& \nonumber K_F (x,x' ; \tau)_{FS} = \frac{1}{(4 \pi 
\hbar \alpha \tau)^2}  \\
\label{propagator_crossed_field_cg} & \times \exp\left(-\frac{i}{\hbar 
}\frac{(x-x')^2}{4 \alpha 
\tau}+\frac{i \lambda^2 \tau}{48 \, \hbar \alpha}(x-x')F^2(x-x')-\frac{i  m^2
c^2 \alpha \tau}{\hbar} \right) \, 
\eal 
which corresponds to the vector potential in the classical path gauge
\be
\mathcal{A}_\mu (\mathcal{P}_c,x) = - \frac{1}{2} \left(F + \frac{\lambda 
\tau}{3} 
F^2 \right)_{\mu \nu}
(x-x')^\nu \, .
\ee

For comparison, in the Fock-Schwinger gauge, the vector potential 
\be
\mathcal{A}_\mu (\mathcal{P}_{FS},x) = - \frac{1}{2} F_{\mu\nu} (x-x')^\nu
\ee
is obtained via the gauge function
\be
\chi = \frac{\lambda \tau}{12} (x-x')^\mu {F^2}_{\mu \nu} (x-x')^\nu \, .
\ee
In terms of the gauge paths, the above gauge function corresponds to the flux through the area bounded by the classical path (\ref{classical_path_em}) and the Fock-Schwinger path which is a
straight line 
\be
\mathcal{P}_{FS} : \, y^\mu (\sigma) = \sigma x^\mu + (1-\sigma) x'^\mu
\ee
with the boundary conditions $y^\mu(1)=x^\mu$, $y^\mu(0)=x'^\mu$.


\subsection{Plane wave case: Volkov Propagator} \label{Volkov}

Another case where the quasiclassical approximation yields the exact result is the interaction of a
charged particle with a plane electromagnetic  wave 
\cite{Volkov_1935,ReissEberly,EberlyReiss,Boudjedaa_1992,Barducci_2003}. The
corresponding propagator for this case is called Volkov propagator. 
The field strength tensor $F_{\mu \nu}$ of a plane wave can be written as
\be
\label{plane_wave_tensor}
F_{\mu \nu} (\phi) = \epsilon_{\mu \nu} f'(\phi)
\ee
where the phase of the wave is defined as $\phi = k x$ and the antisymmetric 
tensor is $\epsilon_{\mu \nu} = k_\mu \epsilon_\nu - k_\nu \epsilon_\mu $ with 
the propagation and polarization  directions $k_\mu$ and $\epsilon_\mu$, 
respectively.

The classical trajectory $\mathcal{P}_c$  via 
Eq.~(\ref{effective_lorentz_force}) for a plane wave reads:
\bal
& k y_c(\sigma) =  \frac{\sigma}{\tau} k(x-x')+kx' \, , \\
& \epsilon y_c(\sigma)  =  \frac{\sigma}{\tau} \epsilon(x-x')+\epsilon x' + 
\frac{\lambda }{\tau }
\left( \tau g_1 (\sigma) -\sigma g_1 (\tau) \right)\, , \\
& \overline{\epsilon} y_c(\sigma)  =  \frac{\sigma}{\tau}
\overline{\epsilon}(x-x')+\overline{\epsilon}x' \, , \\
& \nonumber \overline{k} y_c(\sigma)  = \frac{\lambda \left( \tau
g_1 (\sigma) -\sigma g_1 (\tau) \right)}{k(x-x') \tau} \left(\epsilon(x-x')- 
\lambda \,  g_1 (\tau) \right)   \\
 &+ \frac{\lambda^2}{2\, k(x-x')}\left( \tau g_2 (\sigma) -\sigma g_2 (\tau) 
\right) + \frac{\sigma}{\tau} \overline{k}(x-x')+\overline{k}x'\, ,
\eal
where $\displaystyle g_n (\sigma) = \int_0^\sigma f(k y_c)^n d\sigma'$ and the 
basis $k^\mu, \, \epsilon^\mu,\,\overline{\epsilon}^\mu, \, \overline{k}^\mu $ 
is introduced such that they satisfy $k^2 = \overline{k}^2 =
\epsilon k = \overline{\epsilon} k =  \epsilon \overline{k} =
\overline{\epsilon} \overline{k} = \epsilon \overline{\epsilon} = 0$, $\epsilon^2 =
\overline{\epsilon}^2 = -1 
$, and $k \overline{k} = 1$.

Furthermore, the classical action in the classical path gauge can be written in terms of the new
basis as
\be
S_c = m^2 c^2 \alpha \tau + \int_0^\tau d \sigma \, \frac{1}{4 \alpha} 
\left(2 k \dot{y}_c \, \overline{k}
\dot{y}_c -\epsilon \dot{y}_c^2 -\overline{\epsilon} \dot{y}_c^2 \right) \, .
\ee

Consequently, the Feynman Kernel in the classical path gauge yields
\begin{widetext}
\be
K_F (x,x' ; \tau) =  \frac{1}{(4 \pi \hbar \alpha \tau)^2} \exp\left( - 
\frac{i}{\hbar}\frac{(x-x')^2}{4 \alpha \tau} 
- \frac{i m^2 c^2 \alpha \tau }{\hbar} - \frac{i \lambda^2}{4 \hbar \alpha 
\tau} \left( 
\int_0^\tau 
d \sigma \, f(k y_c)  \right)^2 + 
\frac{i \lambda^2}{4 \hbar \alpha} \int_0^\tau d \sigma \, f(k y_c)^2  \right) 
\, .
\ee
\end{widetext}
The result is more compact due to the absence of the interaction term, see for 
instance Eq.~(31) of \cite{Boudjedaa_1992}. For a constant and uniform plane 
wave one naturally covers Eq.~(\ref{propagator_crossed_field_cg}) and 
in the absence of the field one obtains the relativistic propagator for a free 
particle, see Eq.~(19.28) of \cite{Kleinert}. Furthermore, the Volkov 
propagator can be written in an arbitrary gauge via the 
transformation~(\ref{gauge_trans_kernel}).

\subsection{Plane wave combined with a constant and uniform electromagnetic field case}

In the last example we will obtain the relativistic propagator for a charged particle interacting
with an arbitrary plane wave combined with a constant and uniform electromagnetic field. 

The associated field strength tensor can be written as
\be
F^{\mu \nu} (\phi) = {F_0}^{\mu \nu} + f^{\mu \nu} (\phi)
\ee
where ${F_0}^{\mu \nu}$ and $f^{\mu \nu} (\phi)$ are the field strength tensors of the constant and
uniform electromagnetic field and the plane wave (\ref{plane_wave_tensor}), 
respectively.

Before calculating the classical action, note that since the action is a Lorentz scalar, one
can calculate it in an arbitrary frame of reference. In fact, in an arbitrary reference frame,
there are two fundamental Lorentz invariants which have to be satisfied by any field strength tensor
\bal
 F_{\mu \nu} F^{\mu \nu} &= 2 \, (\vec{B}^2 - \vec{E}^2) \, , \\
 G_{\mu \nu} F^{\mu \nu} &= -4 \, \vec{E} \cdot \vec{B}
\eal
with the dual of the field strength tensor $G_{\mu \nu} = \epsilon^{\mu \nu \alpha \beta} F_{\alpha
\beta} /2 $. Hence, for the constant and uniform electromagnetic field there 
exists such a reference frame that the magnetic field and the electric field can 
be parallel to each other. Furthermore, the direction of the parallel 
magnetic and electric fields can be chosen along the propagation 
direction of the plane wave \cite{Bagrov}. As a consequence, the field strength
tensor of the constant and uniform electromagnetic field ${F_0}^{\mu \nu}$ can be written as
\be
{F_0}^{\mu \nu} = E_0 \, (k^\mu \overline{k}^\nu - k^\nu \overline{k}^\mu) - i B_0 \, 
( \epsilon_{+}^\mu  \epsilon_{-}^\nu - \epsilon_{+}^\nu  \epsilon_{-}^\mu ) \, ,
\ee
where $E_0$ and $B_0$ are the electric field and the magnetic field in 
aforementioned frame, respectively, and the new basis ${\epsilon_{\pm}}^\mu = 
\frac{1}{\sqrt{2}}(\epsilon  \pm i \overline{\epsilon})^\mu$ satisfy 
$\epsilon_{+} \epsilon_{-} = -1 $, $\epsilon_{\pm}^2 = 0$.
Moreover, one can recover the field strength tensor as
\bal
E_0 &= \frac{1}{2}\sqrt{\sqrt{I_1^2+I_2^2}-I_1} \, , \\
B_0 &= -\frac{1}{2}\sqrt{\sqrt{I_1^2+I_2^2}+I_1}
\eal
with $I_1 = {F_0}_{\mu \nu} {F_0}^{\mu \nu}$ and $I_2={G_0}_{\mu \nu} {F_0}^{\mu \nu}$.

Then in the frame of reference where the electric field and the magnetic field of $F_0^{\mu \nu}$
and the propagation direction of $f^{\mu \nu} (\phi)$ are all parallel to each other, the equations
of motion are governed by
\bal
k \ddot{y}_c(\sigma) & = - \lambda E_0 \, k \dot{y}_c (\sigma) \, , \\
\overline{k} \ddot{y}_c(\sigma) & = \lambda E_0 \, \overline{k} \dot{y}_c (\sigma)  +
\frac{\lambda  \epsilon \dot{y}_c  \, \dot{f} }{ k \dot{y}_c } \, , \\
\epsilon_{+} \ddot{y}_c (\sigma) &= - i \lambda B_0 \, \epsilon_{+} \dot{y}_c (\sigma) + \frac{\lambda
\dot{f}}{\sqrt{2}} \, ,\\
\epsilon_{-} \ddot{y}_c (\sigma) &=  i \lambda B_0 \, \epsilon_{-} \dot{y}_c (\sigma) + \frac{\lambda
\dot{f}}{\sqrt{2}} \, .
\eal 

As a consequence, the Feynman Kernel becomes
\begin{subequations}
\label{feynman_kernel_plane_constant}
\be
K_F (x,x') = \sqrt{\left(\frac{1}{2 \pi i \hbar}\right)^4
\det\left(\frac{\del^2 S_c}{\del
x_\mu \del {x'}_\nu } \right)} \exp\left(- \frac{i}{\hbar} S_c \right),
\ee
where the classical action is
\be
S_c = m^2 c^2 \alpha \tau + \int_0^\tau d \sigma \, \frac{1}{2 \alpha} \left(k 
\dot{y}_c \, \overline{k}
\dot{y}_c - \epsilon_{+} \dot{y}_c  \, \epsilon_{-} \dot{y}_c \right)
\ee
with the following solutions of the equations of motion
\begin{widetext}
\bal 
\label{ky}
& k y_c(\sigma)  = \frac{e^{-\lambda E_0 \sigma}-1}{e^{-\lambda E_0 \tau}-1}k(x-x')+kx' \, , \\
\label{kby}
& \epsilon_{+} y_c (\sigma) = \frac{e^{- i \lambda B_0 \sigma}-1}{e^{- i \lambda B_0 \tau}-1} \left[
\epsilon_{+}(x-x') 
- \frac{i}{\sqrt{2}B_0} \left( \frac{e^{i \lambda B_0 (\sigma-\tau)}-1}{1-e^{i \lambda B_0 \sigma}} 
\int_0^\sigma (1- e^{i \lambda B_0 \rho}) \dot{f} d \rho + 
\int_\tau^\sigma  (1- e^{i \lambda B_0 (\rho-\tau)}) \dot{f} d \rho \right) \right] +\epsilon_{+}x' \, , \\
\label{epy}
& \epsilon_{-} y_c (\sigma) = \frac{e^{i \lambda B_0 \sigma}-1}{e^{i \lambda B_0 \tau}-1} \left[
\epsilon_{-}(x-x') 
+ \frac{i}{\sqrt{2}B_0} \left( \frac{e^{-i  \lambda B_0 (\sigma-\tau)}-1}{1-e^{-i  \lambda B_0 \sigma}} 
\int_0^\sigma  (1- e^{-i  \lambda B_0 \rho}) \dot{f} d \rho + 
\int_\tau^\sigma  (1- e^{-i  \lambda B_0 (\rho-\tau)}) \dot{f} d \rho \right) \right] +\epsilon_{-}x' \, , \\
\label{emy}
& \overline{k} y_c (\sigma) = \frac{e^{ \lambda E_0 \sigma}-1}{e^{\lambda E_0 \tau}-1} \left[
\overline{k}(x-x') 
- \frac{1}{E_0} \left( \frac{e^{ \lambda E_0 \tau}-e^{ \lambda E_0 \sigma}}{e^{ \lambda E_0 \sigma}-1} 
\int_0^\sigma  (1- e^{-\lambda E_0 \, \rho}) \frac{\epsilon \dot{y}_c \, \dot{f}}{k \dot{y}_c} d \rho + 
\int_\tau^\sigma (1- e^{- \lambda E_0 (\rho-\tau)}) \frac{\epsilon \dot{y}_c \, \dot{f}}{k \dot{y}_c}  
d \rho \right) \right] + \overline{k}x' \, .
\eal
\end{widetext}
\end{subequations}
Although the closed expression for the Feynman Kernel is very cumbersome and is 
not shown here, it is derived by straightforward calculation
when the plane wave function $f'(\phi)$ and the components of the
field strength tensor of the constant and uniform electromagnetic field 
${F_0}^{\mu \nu}$ are known. Moreover, the form of the Feynman Kernel given by 
Eq.~(\ref{feynman_kernel_plane_constant}) provides considerable convenience to 
the numerical calculations. Simpler expressions for the propagator can be 
obtained in a limit $E_0 \rightarrow 0$ ($B_0 \rightarrow 0$), corresponding to 
a plane wave combined with a constant and uniform magnetic (electric) field 
along the propagation direction of the plane wave.

\section{Conclusion} \label{conc}

We have applied the path-dependent formulation of the gauge theory within the 
Feynman path integral formalism of quantum mechanics for a Klein-Gordon particle 
in an external electromagnetic field. The applied formalism points to a specific 
gauge when a significant simplification of the expression of quasiclassical 
propagators is obtained. The simplification  is due to the fact that the 
interaction part of the classical action vanishes in this gauge. In the 
path-dependent formulation the optimal gauge corresponds to choice of the 
classical path in the definition of the vector potential. 

Specifically, we have calculated the quasiclassical propagators of a scalar  
charged particle interacting with an arbitrary constant and uniform 
electromagnetic field, an arbitrary plane wave and, finally, an arbitrary plane 
wave combined with an arbitrary constant and uniform electromagnetic field. It 
is shown that in the classical path gauge the expressions for the quasiclassical 
propagators, which yield the exact result for above configurations, are more 
compact.

\section*{Acknowledgments}

We are grateful to M. Klaiber, S. Meuren, and A. W\"{o}llert for valuable
discussions.

\bibliography{yakaboylu_bibliography}

\begin{thebibliography}{30}%
\makeatletter
\providecommand \@ifxundefined [1]{%
 \@ifx{#1\undefined}
}%
\providecommand \@ifnum [1]{%
 \ifnum #1\expandafter \@firstoftwo
 \else \expandafter \@secondoftwo
 \fi
}%
\providecommand \@ifx [1]{%
 \ifx #1\expandafter \@firstoftwo
 \else \expandafter \@secondoftwo
 \fi
}%
\providecommand \natexlab [1]{#1}%
\providecommand \enquote  [1]{``#1''}%
\providecommand \bibnamefont  [1]{#1}%
\providecommand \bibfnamefont [1]{#1}%
\providecommand \citenamefont [1]{#1}%
\providecommand \href@noop [0]{\@secondoftwo}%
\providecommand \href [0]{\begingroup \@sanitize@url \@href}%
\providecommand \@href[1]{\@@startlink{#1}\@@href}%
\providecommand \@@href[1]{\endgroup#1\@@endlink}%
\providecommand \@sanitize@url [0]{\catcode `\\12\catcode `\$12\catcode
  `\&12\catcode `\#12\catcode `\^12\catcode `\_12\catcode `\%12\relax}%
\providecommand \@@startlink[1]{}%
\providecommand \@@endlink[0]{}%
\providecommand \url  [0]{\begingroup\@sanitize@url \@url }%
\providecommand \@url [1]{\endgroup\@href {#1}{\urlprefix }}%
\providecommand \urlprefix  [0]{URL }%
\providecommand \Eprint [0]{\href }%
\providecommand \doibase [0]{http://dx.doi.org/}%
\providecommand \selectlanguage [0]{\@gobble}%
\providecommand \bibinfo  [0]{\@secondoftwo}%
\providecommand \bibfield  [0]{\@secondoftwo}%
\providecommand \translation [1]{[#1]}%
\providecommand \BibitemOpen [0]{}%
\providecommand \bibitemStop [0]{}%
\providecommand \bibitemNoStop [0]{.\EOS\space}%
\providecommand \EOS [0]{\spacefactor3000\relax}%
\providecommand \BibitemShut  [1]{\csname bibitem#1\endcsname}%
\let\auto@bib@innerbib\@empty
\bibitem [{\citenamefont {Weinberg}(2013)}]{Weinberg_qm}%
  \BibitemOpen
  \bibfield  {author} {\bibinfo {author} {\bibfnamefont {S.}~\bibnamefont
  {Weinberg}},\ }\href@noop {} {\emph {\bibinfo {title} {Lectures on Quantum
  Mechanics}}}\ (\bibinfo  {publisher} {Cambridge University Press},\ \bibinfo
  {address} {New York},\ \bibinfo {year} {2013})\BibitemShut {NoStop}%
\bibitem [{\citenamefont {Landau}\ and\ \citenamefont
  {Lifshitz}(1981)}]{Landau_3}%
  \BibitemOpen
  \bibfield  {author} {\bibinfo {author} {\bibfnamefont {L.~D.}\ \bibnamefont
  {Landau}}\ and\ \bibinfo {author} {\bibfnamefont {E.~M.}\ \bibnamefont
  {Lifshitz}},\ }\href@noop {} {\emph {\bibinfo {title} {Quantum Mechanics}}},\
  \bibinfo {series} {Course of Theoretical Physics}, Vol.~\bibinfo {volume}
  {3}\ (\bibinfo  {publisher} {Butterworth Heinemann},\ \bibinfo {address}
  {Oxford},\ \bibinfo {year} {1981})\BibitemShut {NoStop}%
\bibitem [{\citenamefont {Feynman}(1948)}]{Feynman_1948}%
  \BibitemOpen
  \bibfield  {author} {\bibinfo {author} {\bibfnamefont {R.~P.}\ \bibnamefont
  {Feynman}},\ }\href@noop {} {\bibfield  {journal} {\bibinfo  {journal} {Rev.
  Mod. Phys.}\ }\textbf {\bibinfo {volume} {20}},\ \bibinfo {pages} {367}
  (\bibinfo {year} {1948})}\BibitemShut {NoStop}%
\bibitem [{\citenamefont {Feynman}\ and\ \citenamefont
  {Hibbs}(1965)}]{Feynman_Hibbs}%
  \BibitemOpen
  \bibfield  {author} {\bibinfo {author} {\bibfnamefont {R.~P.}\ \bibnamefont
  {Feynman}}\ and\ \bibinfo {author} {\bibfnamefont {A.~R.}\ \bibnamefont
  {Hibbs}},\ }\href@noop {} {\emph {\bibinfo {title} {Quantum Mechanics and
  Path Integrals}}}\ (\bibinfo  {publisher} {McGraw-Hill},\ \bibinfo {address}
  {New York},\ \bibinfo {year} {1965})\BibitemShut {NoStop}%
\bibitem [{\citenamefont {Schulman}(1981)}]{Schulman}%
  \BibitemOpen
  \bibfield  {author} {\bibinfo {author} {\bibfnamefont {L.~S.}\ \bibnamefont
  {Schulman}},\ }\href@noop {} {\emph {\bibinfo {title} {Techniques and
  Applications of Path Integration}}}\ (\bibinfo  {publisher} {Wiley},\
  \bibinfo {address} {New York},\ \bibinfo {year} {1981})\BibitemShut {NoStop}%
\bibitem [{\citenamefont {Kleinert}(2009)}]{Kleinert}%
  \BibitemOpen
  \bibfield  {author} {\bibinfo {author} {\bibfnamefont {H.}~\bibnamefont
  {Kleinert}},\ }\href@noop {} {\emph {\bibinfo {title} {Path Integrals in
  Quantum Mechanics, Statistics, Polymer Physics, and Financial Markets}}}\
  (\bibinfo  {publisher} {World Scientiﬁc},\ \bibinfo {address} {Singapore},\
  \bibinfo {year} {2009})\BibitemShut {NoStop}%
\bibitem [{\citenamefont {Volkov}(1935)}]{Volkov_1935}%
  \BibitemOpen
  \bibfield  {author} {\bibinfo {author} {\bibfnamefont {D.~M.}\ \bibnamefont
  {Volkov}},\ }\href {\doibase 10.1007/BF01331022} {\bibfield  {journal}
  {\bibinfo  {journal} {Z. Phys.}\ }\textbf {\bibinfo {volume} {94}},\ \bibinfo
  {pages} {250} (\bibinfo {year} {1935})}\BibitemShut {NoStop}%
\bibitem [{\citenamefont {Reiss}\ and\ \citenamefont
  {Eberly}(1966)}]{ReissEberly}%
  \BibitemOpen
  \bibfield  {author} {\bibinfo {author} {\bibfnamefont {H.~R.}\ \bibnamefont
  {Reiss}}\ and\ \bibinfo {author} {\bibfnamefont {J.~H.}\ \bibnamefont
  {Eberly}},\ }\href@noop {} {\bibfield  {journal} {\bibinfo  {journal} {Phys.
  Rev.}\ }\textbf {\bibinfo {volume} {151}},\ \bibinfo {pages} {1058} (\bibinfo
  {year} {1966})}\BibitemShut {NoStop}%
\bibitem [{\citenamefont {Eberly}\ and\ \citenamefont
  {Reiss}(1966)}]{EberlyReiss}%
  \BibitemOpen
  \bibfield  {author} {\bibinfo {author} {\bibfnamefont {J.~H.}\ \bibnamefont
  {Eberly}}\ and\ \bibinfo {author} {\bibfnamefont {H.~R.}\ \bibnamefont
  {Reiss}},\ }\href@noop {} {\bibfield  {journal} {\bibinfo  {journal} {Phys.
  Rev.}\ }\textbf {\bibinfo {volume} {145}},\ \bibinfo {pages} {1035} (\bibinfo
  {year} {1966})}\BibitemShut {NoStop}%
\bibitem [{\citenamefont {Boudjedaa}\ \emph {et~al.}(1992)\citenamefont
  {Boudjedaa}, \citenamefont {Chetouani}, \citenamefont {Guechi},\ and\
  \citenamefont {Hammann}}]{Boudjedaa_1992}%
  \BibitemOpen
  \bibfield  {author} {\bibinfo {author} {\bibfnamefont {T.}~\bibnamefont
  {Boudjedaa}}, \bibinfo {author} {\bibfnamefont {L.}~\bibnamefont
  {Chetouani}}, \bibinfo {author} {\bibfnamefont {L.}~\bibnamefont {Guechi}}, \
  and\ \bibinfo {author} {\bibfnamefont {T.~F.}\ \bibnamefont {Hammann}},\
  }\href@noop {} {\bibfield  {journal} {\bibinfo  {journal} {Phys. Scr. A}\
  }\textbf {\bibinfo {volume} {46}},\ \bibinfo {pages} {289} (\bibinfo {year}
  {1992})}\BibitemShut {NoStop}%
\bibitem [{\citenamefont {Barducci}\ and\ \citenamefont
  {Giachetti}(2003)}]{Barducci_2003}%
  \BibitemOpen
  \bibfield  {author} {\bibinfo {author} {\bibfnamefont {A.}~\bibnamefont
  {Barducci}}\ and\ \bibinfo {author} {\bibfnamefont {R.}~\bibnamefont
  {Giachetti}},\ }\href@noop {} {\bibfield  {journal} {\bibinfo  {journal} {J.
  Phys. A: Math. Gen.}\ }\textbf {\bibinfo {volume} {36}},\ \bibinfo {pages}
  {8129} (\bibinfo {year} {2003})}\BibitemShut {NoStop}%
\bibitem [{\citenamefont {DeWitt}(1962)}]{DeWitt_1962}%
  \BibitemOpen
  \bibfield  {author} {\bibinfo {author} {\bibfnamefont {B.~S.}\ \bibnamefont
  {DeWitt}},\ }\href@noop {} {\bibfield  {journal} {\bibinfo  {journal} {Phys.
  Rev.}\ }\textbf {\bibinfo {volume} {125}},\ \bibinfo {pages} {2189} (\bibinfo
  {year} {1962})}\BibitemShut {NoStop}%
\bibitem [{\citenamefont {Mandelstam}(1962)}]{Mandelstam_1962}%
  \BibitemOpen
  \bibfield  {author} {\bibinfo {author} {\bibfnamefont {S.}~\bibnamefont
  {Mandelstam}},\ }\href@noop {} {\bibfield  {journal} {\bibinfo  {journal}
  {Ann. Phys.}\ }\textbf {\bibinfo {volume} {19}},\ \bibinfo {pages} {1}
  (\bibinfo {year} {1962})}\BibitemShut {NoStop}%
\bibitem [{\citenamefont {Wu}\ and\ \citenamefont {Yang}(1975)}]{Wu_Yang_1975}%
  \BibitemOpen
  \bibfield  {author} {\bibinfo {author} {\bibfnamefont {T.~T.}\ \bibnamefont
  {Wu}}\ and\ \bibinfo {author} {\bibfnamefont {C.~N.}\ \bibnamefont {Yang}},\
  }\href@noop {} {\bibfield  {journal} {\bibinfo  {journal} {Phys. Rev. D}\
  }\textbf {\bibinfo {volume} {12}},\ \bibinfo {pages} {3845} (\bibinfo {year}
  {1975})}\BibitemShut {NoStop}%
\bibitem [{\citenamefont {Yakaboylu}\ and\ \citenamefont
  {Hatsagortsyan}(2013)}]{Yakaboylu_2013}%
  \BibitemOpen
  \bibfield  {author} {\bibinfo {author} {\bibfnamefont {E.}~\bibnamefont
  {Yakaboylu}}\ and\ \bibinfo {author} {\bibfnamefont {K.~Z.}\ \bibnamefont
  {Hatsagortsyan}},\ }\href@noop {} {\enquote {\bibinfo {title} {On the
  quantization of the nonintegrable phase in electrodynamics},}\ } (\bibinfo
  {year} {2013}),\ \Eprint {http://arxiv.org/abs/1309.0715} {arXiv:1309.0715}
  \BibitemShut {NoStop}%
\bibitem [{\citenamefont {Belinfante}(1962)}]{Belinfante_1962}%
  \BibitemOpen
  \bibfield  {author} {\bibinfo {author} {\bibfnamefont {F.~J.}\ \bibnamefont
  {Belinfante}},\ }\href@noop {} {\bibfield  {journal} {\bibinfo  {journal}
  {Phys. Rev.}\ }\textbf {\bibinfo {volume} {128}},\ \bibinfo {pages} {2832}
  (\bibinfo {year} {1962})}\BibitemShut {NoStop}%
\bibitem [{\citenamefont {Rohrlich}\ and\ \citenamefont
  {Strocchi}(1965)}]{Rohrlich_Strocchi_1965}%
  \BibitemOpen
  \bibfield  {author} {\bibinfo {author} {\bibfnamefont {F.}~\bibnamefont
  {Rohrlich}}\ and\ \bibinfo {author} {\bibfnamefont {F.}~\bibnamefont
  {Strocchi}},\ }\href@noop {} {\bibfield  {journal} {\bibinfo  {journal}
  {Phys. Rev.}\ }\textbf {\bibinfo {volume} {139}},\ \bibinfo {pages} {B476}
  (\bibinfo {year} {1965})}\BibitemShut {NoStop}%
\bibitem [{\citenamefont {Fock}(1937)}]{Fock_1937}%
  \BibitemOpen
  \bibfield  {author} {\bibinfo {author} {\bibfnamefont {V.}~\bibnamefont
  {Fock}},\ }\href@noop {} {\bibfield  {journal} {\bibinfo  {journal} {Phys.
  Rev.}\ }\textbf {\bibinfo {volume} {12}},\ \bibinfo {pages} {404} (\bibinfo
  {year} {1937})}\BibitemShut {NoStop}%
\bibitem [{\citenamefont {Feynman}(1950)}]{Feynman_1950}%
  \BibitemOpen
  \bibfield  {author} {\bibinfo {author} {\bibfnamefont {R.}~\bibnamefont
  {Feynman}},\ }\href {http://dx.doi.org/10.1103/PhysRev.80.440} {\bibfield
  {journal} {\bibinfo  {journal} {Physical Review}\ }\textbf {\bibinfo {volume}
  {80}},\ \bibinfo {pages} {440} (\bibinfo {year} {1950})}\BibitemShut
  {NoStop}%
\bibitem [{\citenamefont {Schwinger}(1951)}]{Schwinger_1951}%
  \BibitemOpen
  \bibfield  {author} {\bibinfo {author} {\bibfnamefont {J.}~\bibnamefont
  {Schwinger}},\ }\href@noop {} {\bibfield  {journal} {\bibinfo  {journal}
  {Phys. Rev.}\ }\textbf {\bibinfo {volume} {82}},\ \bibinfo {pages} {664}
  (\bibinfo {year} {1951})}\BibitemShut {NoStop}%
\bibitem [{\citenamefont {Dirac}(1964)}]{Dirac_qm}%
  \BibitemOpen
  \bibfield  {author} {\bibinfo {author} {\bibfnamefont {P.~A.~M.}\
  \bibnamefont {Dirac}},\ }\href@noop {} {\emph {\bibinfo {title} {Lectures on
  Quantum Mechanics}}}\ (\bibinfo  {publisher} {Yeshiva University},\ \bibinfo
  {address} {New York},\ \bibinfo {year} {1964})\BibitemShut {NoStop}%
\bibitem [{\citenamefont {Fradkin}\ and\ \citenamefont
  {Gitman}(1991)}]{Fradkin_1991}%
  \BibitemOpen
  \bibfield  {author} {\bibinfo {author} {\bibfnamefont {E.~S.}\ \bibnamefont
  {Fradkin}}\ and\ \bibinfo {author} {\bibfnamefont {D.~M.}\ \bibnamefont
  {Gitman}},\ }\href@noop {} {\bibfield  {journal} {\bibinfo  {journal} {Phys.
  Rev. D}\ }\textbf {\bibinfo {volume} {44}},\ \bibinfo {pages} {3230}
  (\bibinfo {year} {1991})}\BibitemShut {NoStop}%
\bibitem [{\citenamefont {Brink}\ \emph {et~al.}(1977)\citenamefont {Brink},
  \citenamefont {Vecchia},\ and\ \citenamefont {Howe}}]{Brink_1977}%
  \BibitemOpen
  \bibfield  {author} {\bibinfo {author} {\bibfnamefont {L.}~\bibnamefont
  {Brink}}, \bibinfo {author} {\bibfnamefont {P.~D.}\ \bibnamefont {Vecchia}},
  \ and\ \bibinfo {author} {\bibfnamefont {P.}~\bibnamefont {Howe}},\
  }\href@noop {} {\bibfield  {journal} {\bibinfo  {journal} {Nucl. Phys. B}\
  }\textbf {\bibinfo {volume} {118}},\ \bibinfo {pages} {76} (\bibinfo {year}
  {1977})}\BibitemShut {NoStop}%
\bibitem [{\citenamefont {Polyakov}(1987)}]{Polyakov_st}%
  \BibitemOpen
  \bibfield  {author} {\bibinfo {author} {\bibfnamefont {A.~M.}\ \bibnamefont
  {Polyakov}},\ }\href@noop {} {\emph {\bibinfo {title} {Gauge Fields and
  Strings}}}\ (\bibinfo  {publisher} {Harwood Academic},\ \bibinfo {address}
  {Chur, Switzerland},\ \bibinfo {year} {1987})\BibitemShut {NoStop}%
\bibitem [{\citenamefont {VanVleck}(1928)}]{VanVleck_1928}%
  \BibitemOpen
  \bibfield  {author} {\bibinfo {author} {\bibfnamefont {J.~H.}\ \bibnamefont
  {VanVleck}},\ }\href@noop {} {\bibfield  {journal} {\bibinfo  {journal}
  {Proc. Nat. Acad. Sci. U.S.A.}\ }\textbf {\bibinfo {volume} {14}},\ \bibinfo
  {pages} {178} (\bibinfo {year} {1928})}\BibitemShut {NoStop}%
\bibitem [{\citenamefont {Morette}(1951)}]{Morette_1951}%
  \BibitemOpen
  \bibfield  {author} {\bibinfo {author} {\bibfnamefont {C.}~\bibnamefont
  {Morette}},\ }\href@noop {} {\bibfield  {journal} {\bibinfo  {journal} {Phys.
  Rev.}\ }\textbf {\bibinfo {volume} {81}},\ \bibinfo {pages} {848} (\bibinfo
  {year} {1951})}\BibitemShut {NoStop}%
\bibitem [{\citenamefont {DeWitt}(1957)}]{DeWitt_1957}%
  \BibitemOpen
  \bibfield  {author} {\bibinfo {author} {\bibfnamefont {B.~S.}\ \bibnamefont
  {DeWitt}},\ }\href@noop {} {\bibfield  {journal} {\bibinfo  {journal} {Rev.
  Mod. Phys.}\ }\textbf {\bibinfo {volume} {29}},\ \bibinfo {pages} {377}
  (\bibinfo {year} {1957})}\BibitemShut {NoStop}%
\bibitem [{\citenamefont {Pauli}(1973)}]{Pauli_1973}%
  \BibitemOpen
  \bibfield  {author} {\bibinfo {author} {\bibfnamefont {W.}~\bibnamefont
  {Pauli}},\ }\href@noop {} {\emph {\bibinfo {title} {Lectures on Physics}}}\
  (\bibinfo  {publisher} {MIT Press},\ \bibinfo {year} {1973})\BibitemShut
  {NoStop}%
\bibitem [{\citenamefont {Zwiebach}(2004)}]{Zwiebach_st}%
  \BibitemOpen
  \bibfield  {author} {\bibinfo {author} {\bibfnamefont {B.}~\bibnamefont
  {Zwiebach}},\ }\href@noop {} {\emph {\bibinfo {title} {A First Course in
  String Theory}}}\ (\bibinfo  {publisher} {Cambridge University Press},\
  \bibinfo {address} {New York},\ \bibinfo {year} {2004})\BibitemShut {NoStop}%
\bibitem [{\citenamefont {Bagrov}\ and\ \citenamefont {Gitman}(1990)}]{Bagrov}%
  \BibitemOpen
  \bibfield  {author} {\bibinfo {author} {\bibfnamefont {V.~G.}\ \bibnamefont
  {Bagrov}}\ and\ \bibinfo {author} {\bibfnamefont {D.~M.}\ \bibnamefont
  {Gitman}},\ }\href@noop {} {\emph {\bibinfo {title} {Exact Solutions of
  Relativistic Wave Equations, Sec.~16}}}\ (\bibinfo  {publisher} {Kluwer
  Academic Publishers},\ \bibinfo {address} {Dordrecht, Boston, London},\
  \bibinfo {year} {1990})\BibitemShut {NoStop}%
\end{thebibliography}%

\end{document}